\documentclass[11pt]{article}

\usepackage{epsfig,subfigure}
\usepackage{amssymb, amsmath}
\usepackage{color}
\usepackage{hyperref}
\usepackage{graphicx}
\usepackage{color}

\usepackage{amssymb}
\usepackage{amsmath,amsfonts,amssymb}

\usepackage{verbatim}
\usepackage{stfloats}
\usepackage[bookmarks=false]{}

\newtheorem{theorem}{Theorem}

\newtheorem{definition}{Definition}
\newtheorem{lemma}{Lemma}

\date{}

\begin{document}
\title{Generalized Degrees of Freedom of the Symmetric $K$-User Interference Channel under Finite Precision CSIT}
\author{ \normalsize Arash Gholami Davoodi and Syed A. Jafar \\
{\small Center for Pervasive Communications and Computing (CPCC)}\\
{\small University of California Irvine, Irvine, CA 92697}\\
{\small \it Email: \{gholamid, syed\}@uci.edu}
}
\maketitle

\begin{abstract}
The  generalized degrees of freedom (GDoF)  characterization of the symmetric $K$-user interference channel is obtained under finite precision channel state information at the transmitters (CSIT). The symmetric setting is where each cross channel is capable of carrying $\alpha$ degrees of freedom (DoF) while each direct channel is capable of carrying $1$ DoF. Remarkably, under finite precision CSIT the symmetric $K$-user interference channel loses all the GDoF benefits of interference alignment. The GDoF per user diminish with the number of users everywhere except in the very strong (optimal for every receiver to decode all messages) and very weak (optimal to treat all interference as noise) interference regimes. The result stands in sharp contrast to prior work on the symmetric setting under perfect CSIT, where the GDoF per user remain undiminished  due to interference alignment. The result also stands in contrast to prior work on a subclass of asymmetric settings under finite precision CSIT, i.e., the topological interference management problem, where  interference alignment plays a crucial role and provides substantial GDoF benefits. 

\end{abstract}
\newpage

\section{Introduction}
The capacity of the $K$ user wireless interference channel is one of the holy grails of network information theory. Much progress has been made on this problem recently through the pursuit of capacity approximations \cite{ADT_FnT}. In particular, the studies of the degrees of freedom (DoF) of interference networks have produced some of the most intriguing results \cite{Jafar_FnT}. On the one hand, if the channel state information at the transmitters (CSIT) is perfect, then ``everyone gets half the cake" \cite{Cadambe_Jafar_int, Nazer_Gastpar_Jafar_Vishwanath, Motahari_Gharan_Khandani,Wu_Shamai_Verdu}, thus circumventing the interference barrier \cite{tse_mobicom}. On the other hand, if the CSIT is restricted to finite precision \cite{Lapidoth_Shamai_Wigger_BC},  then it is shown that the DoF collapse entirely \cite{Arash_Jafar_GC14}, rendering the interference barrier fundamentally insurmountable. The  divergent conclusions representing  optimistic and pessimistic extremes are attributable to the limitations of the underlying assumptions. First, perfect CSIT is problematic as an overly optimistic assumption that opens the door to powerful but fragile interference alignment schemes. This limitation may be avoided by relying only on finite precision CSIT as a robust alternative. Second, the DoF metric by itself is also problematic as it enforces an overly pessimistic perspective, where all channels, whether desired or undesired, are of comparable strength (each capable of carrying 1 DoF), thus relinquishing the benefits of robust interference management schemes that are capable of optimally dealing with weak or strong interference. This limitation is avoided by relying instead on the Generalized Degrees of Freedom (GDoF) metric \cite{Etkin_Tse_Wang}, intended to capture the diversity of channel strengths. A GDoF characterization is also the next step after DoF characterizations in the progressive refinements approach to capacity approximations. Thus, the search for fundamental limits of robust interference management motivates GDoF studies under finite precision CSIT.

In spite of its relevance, until recently there has been very little progress on GDoF characterizations under finite precision CSIT. This is mainly because conventional arguments have repeatedly failed to produce useful outer bounds in this setting. Recall that even the collapse of DoF under finite precision CSIT, as conjectured by Lapidoth, Shamai and Wigger in 2005 \cite{Lapidoth_Shamai_Wigger_BC}, and featured at the ``Open Problems" session at the inaugural ITA workshop in 2006, remained unresolved for nearly a decade. All available outer bounds either implicitly allowed too much channel knowledge so that the DoF could not collapse, or showed a collapse of DoF under degenerate extremes -- where the channel directions are assumed statistically indistinguishable (e.g., isotropic \cite{Jafar_scalar, Huang_Jafar_Shamai_Vishwanath, Zhu_Guo_MIMOIC, Varanasi_noCSIT}). The original attempt by Lapidoth, Shamai and Wigger \cite{Lapidoth_Shamai_Wigger_BC} used the Csiszar sum identity.  Weingarten, Shamai and Kramer \cite{Weingarten_Shamai_Kramer} used a compound channel argument to strengthen the conjecture, but the strengthened conjecture was shown to be false by Gou, Jafar and Wang in \cite{Gou_Jafar_Wang}, and by Maddah-Ali in \cite{Maddah_Compound}. Tandon, Jafar, Shamai and Poor \cite{Tandon_Jafar_Shamai_Poor} considered a stronger setting where perfect channel knowledge is available for some of the users, but here also the conjecture remained unresolved.  Another attempt at proving these conjectures, made in \cite{Hao_Rasouli_Clerckx} based on the extremal inequality of \cite{Liu_Viswanath},   was also unsuccessful. 

The conjectured collapse of DoF under finite precision CSIT was ultimately established recently  in \cite{Arash_Jafar_GC14}. Reminiscent of Korner and Marton's work on the images of a set in \cite{Korner_Marton_images}, the approach of \cite{Arash_Jafar_GC14} is based on estimating the difference in the size of the images of the set of codewords as seen by different users. The  essence of the problem is the need to align signals at one receiver while keeping them separate at another.  Each set of resolvable images at the desired receiver, which corresponds to an aligned (unresolvable) image at the undesired receiver, is known as an ``Aligned Image Set" \cite{Arash_Jafar_GC14}. The average size of this set is the key determinant of the degree to which signals can be aligned. The  approach of \cite{Arash_Jafar_GC14}, in short the AIS approach, is based on a combinatorial accounting of the average size of the aligned image sets to produce a new set of ``robust" information theoretic outer bounds, i.e., bounds that are useful under finite precision CSIT.

By expanding on the AIS approach of \cite{Arash_Jafar_GC14}, there has been recent progress in GDoF characterizations under finite precision CSIT for some scenarios of interest, especially in the 2 user setting \cite{Arash_Jafar_GC15}. For the 2 user interference channel, as noted in \cite{Arash_Jafar_GC15},  the GDoF under finite precision CSIT are the same as the GDoF under perfect CSIT \cite{Etkin_Tse_Wang}. This is because the optimal achievable schemes for $K=2$ are inherently robust, requiring neither zero forcing nor interference alignment. Remarkably, \cite{Arash_Jafar_GC15} also characterizes the GDoF under finite precision CSIT of the 2 user X channel where each of the two transmitters has an independent message for each of the two receivers, as well as the GDoF of the MISO BC formed by allowing full cooperation between the two transmitters. Under perfect CSIT, the X channel benefits from interference alignment \cite{Jafar_Shamai,Huang_Cadambe_Jafar,Niesen_Maddah_Ali_X} while the MISO BC relies on zero forcing. However, under finite precision CSIT, \cite{Arash_Jafar_GC15} shows that both of these benefits are lost. The X channel loses the GDoF gains of interference alignment and reduces to its underlying interference channels. The MISO BC relinquishes the GDoF gains from zero-forcing and retains only the gains from interference enhancement \cite{Maric_Dabora_Goldsmith,Joudeh_Clerckx} that is enabled by the cooperation between the two transmit antennas. 

For the $K$ user interference channel, the GDoF under finite precision CSIT are available implicitly for two regimes of interest, based on separate works focusing on interference that is weak enough to be optimally treated as white noise \cite{Geng_TIN_opt} and on interference that is strong enough to be comparable to the desired signal \cite{Jafar_TIM,Naderi_Avestimehr}. For weak interference, the GDoF optimality of treating interference as noise (TIN) is studied in \cite{Geng_TIN_opt}, where it is shown that  TIN is GDoF optimal  in a $K$ user interference channel if for every user, the desired channel is at least strong as the sum of the strengths of the strongest cross channel to and from that user, with all strengths measured in dB scale. While originally shown under perfect CSIT, the robustness of TIN guarantees that the result also holds under finite precision CSIT. Strong interference, i.e., interference that can be as strong as the desired signals, prompts the interference avoidance question, which is given a fundamental formulation in the topological interference management (TIM) framework in \cite{Jafar_TIM}. Originally presented as a study of DoF of a partially connected network with no CSIT beyond the connectivity, TIM may be also interpreted as a GDoF study under finite precision CSIT. Surprisingly, interference alignment plays a crucial role in the TIM problem, even though the CSIT is restricted to finite precision. Standing in contrast with the 2 user $X$ channel setting where all GDoF benefits of interference alignment are lost under finite precision CSIT \cite{Arash_Jafar_GC15}, the re-emergence of interference alignment as a critical ingredient for TIM with $K$ users \cite{Jafar_TIM} underscores how new aspects of the problem are revealed only as the number of users increases. In particular it motivates  further study of the $K$ user setting. 

A GDoF study for the $K$ user setting in its full generality would involve $O(K^2)$ channel strength parameters, giving rise to potentially an exponential number of distinct regimes. To avoid the immediate curse of dimensionality, a promising approach is to focus initially on the most interesting, albeit narrower perspectives, in order to accumulate new insights which may eventually serve as the building blocks for a comprehensive solution as the understanding of the problem matures. TIN  \cite{Geng_TIN_opt} and TIM \cite{Jafar_TIM}, as described above, offer two such perspectives by addressing weak and strong interference, respectively. But what about  intermediate regimes, e.g., where interference is neither weak enough to be optimally treated as noise nor strong enough to be comparable to the desired signal? Motivated by this question, in this work we initiate the study the GDoF under finite precision CSIT for the entire range of cross channel strengths. However, in order to avoid the explosion of parameters,  we  focus on the decidedly narrow but interesting perspective of the symmetric setting where all direct channels are capable of carrying 1 DoF, and all cross-channels are capable of carrying $\alpha$ DoF. 

\begin{figure}[!h]
\begin{minipage}[c]{0.61\textwidth}
\includegraphics[width=\linewidth]{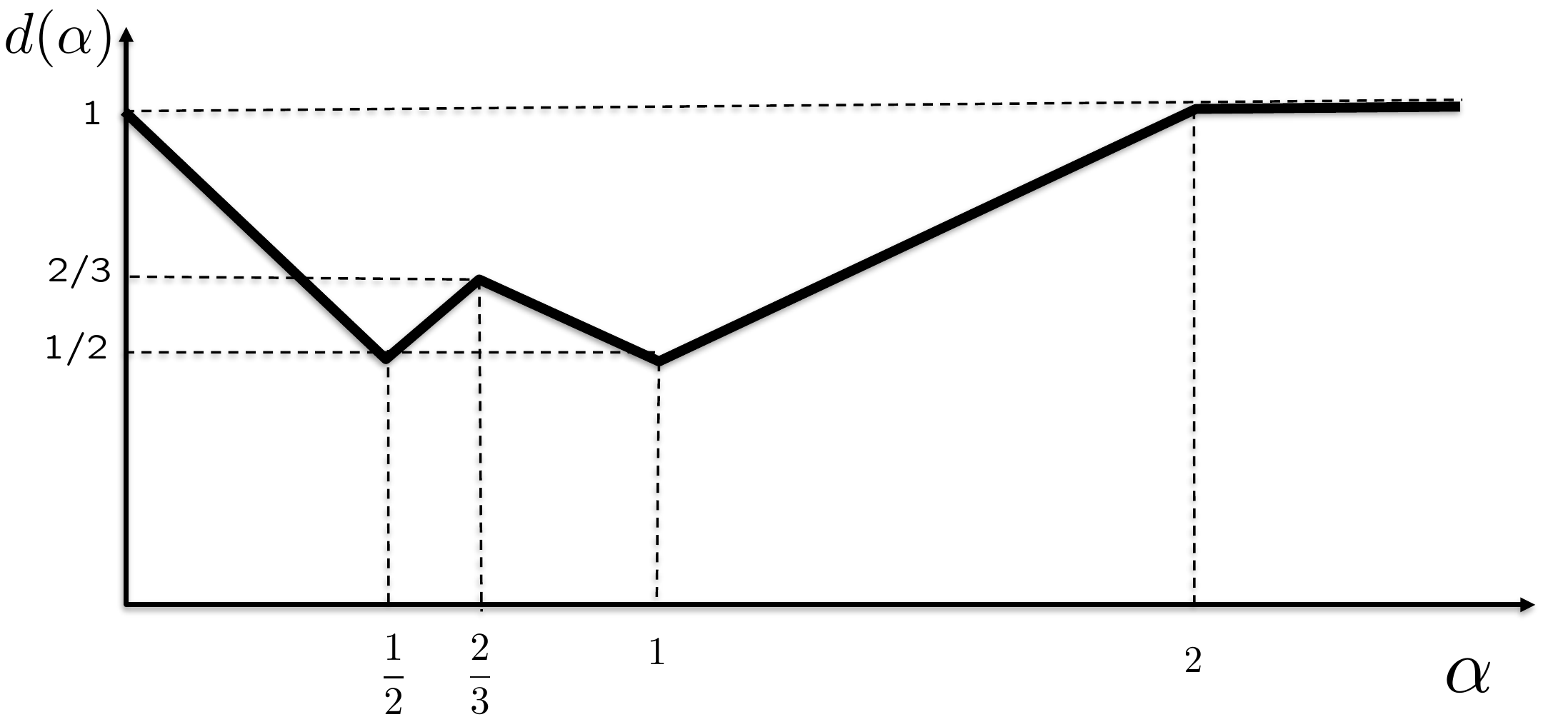}
\end{minipage}~~
\begin{minipage}[c]{0.3\textwidth}
\begin{eqnarray*}
d(\alpha) = \left\{
\begin{array}{ll} 
1-\alpha,&0\leq\alpha\leq \frac{1}{2}\vspace{0.03in}\\ 
\alpha, &\frac{1}{2}<\alpha\leq\frac{2}{3}\vspace{0.03in}\\ 
1-\frac{\alpha}{2},&\frac{2}{3}<\alpha\leq 1\\
\frac{\alpha}{2},& 1<\alpha\leq 2\\
1, &2<\alpha
\end{array}
\right.
\end{eqnarray*}
\end{minipage}
\caption{GDoF/user of the Symmetric $K$-User Interference Channel with Perfect CSIT.}\label{fig:2user}
\end{figure}

The baseline for comparison is the familiar ``W" curve of \cite{Etkin_Tse_Wang} illustrated in Figure \ref{fig:2user}, which depicts the GDoF per user of the symmetric 2 user interference channel as a function of the interference strength $\alpha$. The GDoF curve is piecewise linear, comprised of 5 linear segments, which correspond to very weak, weak, moderate, strong and very strong interference. Under perfect CSIT, the same curve also represents the GDoF per user for the symmetric $K$ user interference channel \cite{Jafar_Vishwanath_GDoF}, i.e., the GDoF per user remain undiminished (due to interference alignment) even as the number of users increases. In this symmetric setting the regime where TIN is GDoF optimal corresponds to $\alpha\leq 1/2$. Note that the DoF value is recovered as a special case of the GDoF corresponding to $\alpha=1$. Also note that when $\alpha=1$, we have $d(\alpha)=1/2$, i.e., everyone gets half the cake.

\begin{figure}[!h]
\begin{minipage}[c]{0.54\textwidth}
	\centerline{\includegraphics[width=4in]{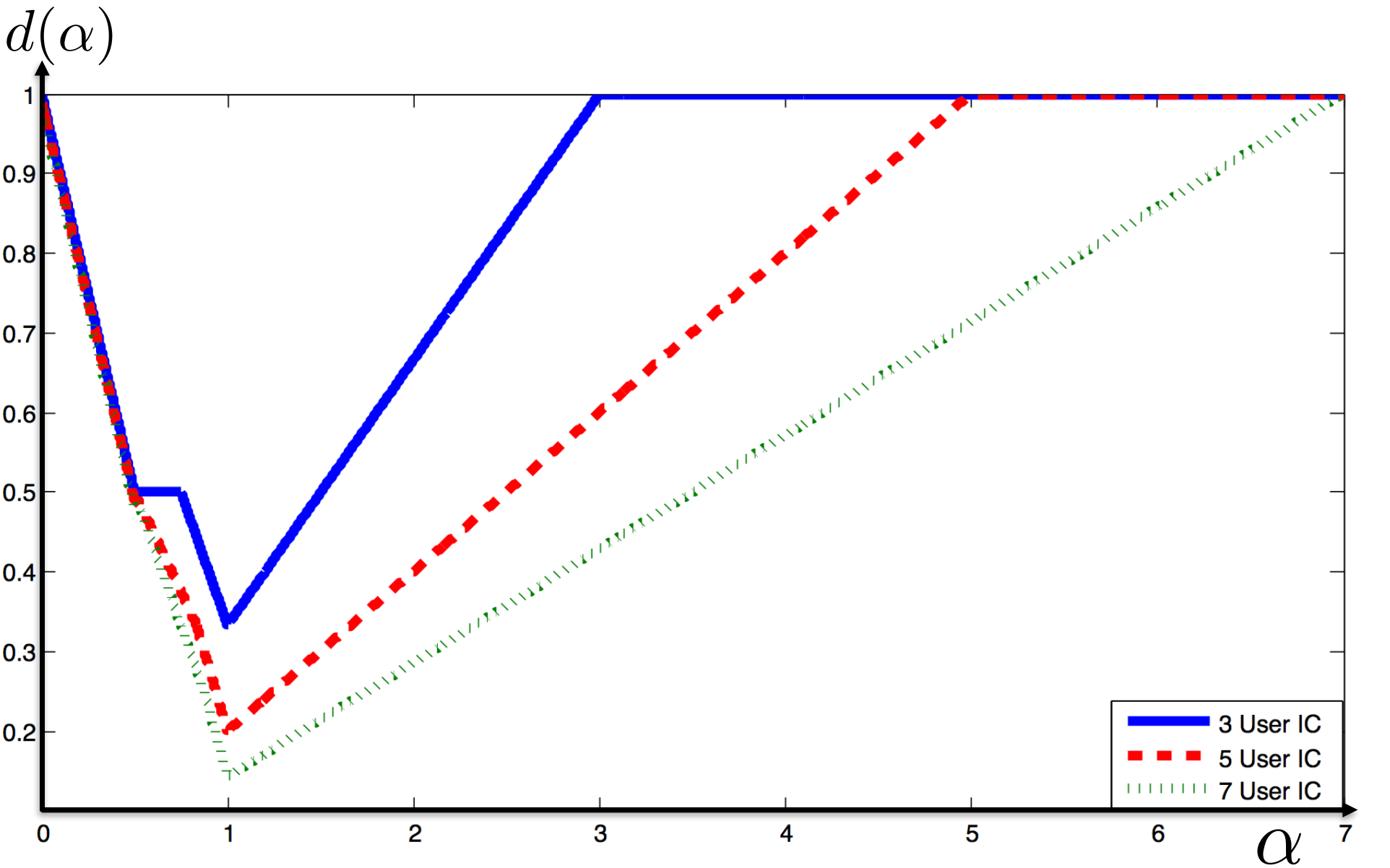}}
\end{minipage}~~~~~
\begin{minipage}[c]{0.35\textwidth}
\begin{eqnarray*}
d(\alpha) = \left\{
\begin{array}{ll} 
1-\alpha,&0\leq\alpha\leq \frac{1}{2}\vspace{0.03in}\\ 
\frac{K-2-(K-3)\alpha}{K-1}, &\frac{1}{2}<\alpha\leq\frac{K}{K+1}\vspace{0.03in}\\ 
1-\left(\frac{K-1}{K}\right)\alpha,&\frac{K}{K+1}<\alpha\leq 1\\
\frac{\alpha}{K},& 1<\alpha\leq K\\
1, &K<\alpha
\end{array}
\right.
\end{eqnarray*}
\end{minipage}
\caption{GDoF/user of the Symmetric $K$-User Interference Channel with Finite Precision CSIT.}\label{fig:Kuser}
\end{figure}

The main result of this work, the GDoF per user of the symmetric $K$ user interference channel under finite precision CSIT, is shown next in Figure \ref{fig:Kuser}. For $K=2$ the GDoF curve under finite precision CSIT is the same as with perfect CSIT. For arbitrary $K$, the GDoF curve remains piecewise linear, comprised of 5 linear segments as before. The GDoF under finite precision CSIT are the same as with perfect CSIT in the very weak interference regime  ($\alpha\leq 1/2$, where it is optimal to treat interference as noise) and the very strong interference regime  ($\alpha > K$, where it is optimal for each receiver to decode all messages).  This is because treating interference as noise, and decoding all messages (multiple access) are robust schemes that do not require infinitely precise CSIT. On the other hand, everywhere else, we note that there is a GDoF loss under finite precision CSIT relative to perfect CSIT. In fact the GDoF per user are a strictly decreasing function of the number of users. These are the regimes where interference alignment provides GDoF benefits under perfect CSIT. Indeed, for the symmetric $K$ user interference channel, the GDoF benefits of interference alignment are lost under finite precision CSIT.  This  is perhaps surprising because it stands in contrast to the studies of TIM which show  that interference alignment is most useful for GDoF optimal schemes even under finite precision CSIT. Evidently, while interference alignment is important for GDoF even with finite precision CSIT, it is not required for the narrower class of symmetric settings studied here. With no need for interference alignment, the GDoF optimal achievable schemes are based on rate splitting and superposition encoding at the transmitters, while each (sub)message is either intended to be decoded or treated as white noise at each receiver. The technical highlight of this work are the tight GDoF outer bounds, which expand upon the AIS approach of \cite{Arash_Jafar_GC14, Arash_Jafar_GC15}. The successful generalization of the AIS argument  is by itself an exciting takeaway from this work, as it paves the way for further expansions that will be necessary in the future to characterize the GDoF under asymmetric settings.

{\it Notation:} For $n\in\mathbb{N}$, define the notation $[n]=\{1,2,\cdots,n\}$. The cardinality of a set $A$ is denoted as $|A|$. The notation $X_{1:i}$ stands for $\{X_1, X_2, \cdots, X_i\}$ and $X^{[n]}$ stands for $X(1), X(2), \cdots X(n)$. For sets $A, B$, the notation $A/B$ refers to the set of elements that are in $A$ but not in $B$.

\section{System Model} {\label{sec-sys}}
\subsection{GDoF of Symmetric $K$ User Interference Channel}
 For the purpose of GDoF studies under finite precision CSIT, the channel model for the $K$ user interference channel is defined by the following input-output equations.
\begin{eqnarray}
Y_k(t)=\sum_{l=1}^{K}\sqrt{P^{\alpha_{kl}}}G_{kl}(t)X_l(t)+Z_k(t),~~ \forall k\in[K].
\end{eqnarray}
At the $t^{th}$ channel use, $X_l(t)$ is the symbol sent from Transmitter $l$, normalized so that it is subject to a  unit power constraint, $Y_k(t)$ is the signal received by Receiver $k$, $Z_k(t)$ is the zero mean unit variance additive white Gaussian noise experienced by Receiver $k$, and $G_{kl}(t)$ are the channel fading coefficients between Transmitter $l$ and Receiver $k$ whose values are bounded away from zero and infinity, i.e., there exist constants $\Delta_1, \Delta_2$ such that $0<\Delta_1\leq 
|G_{kl}(t)|\leq\Delta_2<\infty$. The channel model is parameterized by $P$,  a nominal parameter that approaches infinity in the GDoF limit. The channel strength parameters $\alpha_{kl}$ appear as exponents of $P$. Since we focus on the symmetric setting, for all $k,l\in[K]$, let us  set
\begin{eqnarray*}
\alpha_{kl}&=&\left\{
\begin{array}{ll}
\alpha, & k\neq l\\
1, &k=l.
\end{array}
\right.
\end{eqnarray*}
Thus, the direct channels are capable of carrying 1 DoF, while the cross-channels are capable of carrying $\alpha$ DoF. 

Parameterized by $P$, achievable rates $R_i(P)$ and capacity region $\mathcal{C}(P)$ are defined in the standard Shannon theoretic sense. The GDoF region is defined as
\begin{eqnarray}
\mathcal{D}&=&\{(d_1,\cdots,d_K): \nonumber \\
&&\exists (R_1(P),\cdots, R_K(P))\in\mathcal{C}(P) \mbox{ s. t. } d_k=\lim_{P\rightarrow\infty}\frac{R_k(P)}{C_o(P)}, \forall k\in[K]\} \label {region}
\end{eqnarray}
where $C_o(P)$ is the baseline reference capacity of an additive white Gaussian noise channel $Y=X+N$ with transmit power $P$ and unit variance additive white Gaussian noise. For real settings, $C_o(P)=1/2\log(P)+o(\log(P))$ and for complex settings  $C_o(P)=\log(P)+o(\log(P))$.
Since in this work we are focused exclusively on the sum of the GDoF values, $d_1+d_2+\cdots+d_K$, we will refer to it simply as the GDoF.

As in \cite{Arash_Jafar_GC15}, it is worthwhile to remind ourselves that unlike DoF, the scaling with $P$ in the GDoF framework does not correspond to a physical scaling of powers in a given channel, because of the different power scaling exponents $\alpha_{kl}$. Rather, each $P$ value defines a new channel. Intuitively, this class of channels belong together because, normalized by $\log(P)$, they have (approximately) the same capacity, so that a GDoF characterization simultaneously characterizes the capacity of all the channels in this class within a gap of $o(\log(P))$.

%

\subsection{Finite Precision CSIT}
We assume that the transmitters are aware of the $\alpha_{kl}$ values, i.e., the coarse channel strength parameters, but not the precise $G_{kl}$ values. For the $G_{kl}$ the transmiters are only aware of the joint probability density function. For the real setting, define the set of channel coefficient variables $\mathcal{G}=\{G_{kl}(t): t\in\mathbb{N}, k,l\in[K]\}$. For the complex setting, where $G_{kl}(t)=G_{kl,R}(t)+jG_{kl,I}(t)$, define $\mathcal{G}=\{G_{kl,R}(t): t\in\mathbb{N}, k,l\in[K]\}\cup \{G_{kl,I}(t): t\in\mathbb{N}, k,l\in[K]\}$. We assume that the set of channel coefficients $\mathcal{G}$ is available to the transmitters only up to finite precision, which refers to the ``bounded density assumption". The ``Bounded Density Assumption" is defined as follows.

\begin{definition}[Bounded Density Assumption] A set of random variables, $\mathcal{A}$, is said to satisfy the bounded density assumption if 
 there exists a finite positive constant $f_{\max}$, \label{def1}
\begin{eqnarray*}
	0<f_{\max}<\infty
\end{eqnarray*}
such that for all finite cardinality disjoint subsets $\mathcal{A}_1, \mathcal{ A}_2$ of $\mathcal{A}$, 
\begin{eqnarray*}
	\mathcal{ A}_1\subset \mathcal{A}, \mathcal{A}_2\subset \mathcal{A}, \mathcal{A}_1\cap\mathcal{A}_2=\phi, \mathcal|{A}_1|<\infty, \mathcal|\mathcal{ A}_2|<\infty
\end{eqnarray*}
the conditional probability density functions exist and are bounded as follows,
\begin{eqnarray*}
	\forall A_1, A_2, ~~f_{\mathcal{A}_1|\mathcal{A}_2}(A_1|A_2)&\leq&f_{\max}^{|\mathcal{A}_1|}.
\end{eqnarray*}
\end{definition}

As a special case, if all channel realizations are independent, then it suffices that the marginal densities are bounded by $f_{\max}$.

\section{Main Result}
The main result of this work is stated in the following theorem.
\begin{theorem} \label{Theorem1}
	The GDoF per user of the symmetric $K$-user interference channel under finite precision CSIT is expressed as follows.
\begin{eqnarray}
d(\alpha)= \left\{\begin{array}{l} 1-\alpha,\\ \frac{1}{K-1}(K-2-(K-3)\alpha),\\ 1-\frac{K-1}{K}\alpha,\\\frac{\alpha}{K},\\1, \end{array}
\begin{array}{l} ~~~\alpha\leq \frac{1}{2}\\~~~\frac{1}{2}<\alpha\leq \frac{K}{K+1}\\~~~\frac{K}{K+1}<\alpha\leq 1\\~~~1<\alpha\leq K\\~~~K<\alpha \end{array}\right.
\end{eqnarray}
\end{theorem}
%
%
\section{Proof of Theorem \ref{Theorem1}}
\subsection{Outer Bound}
For notational simplicity, let us define
\begin{eqnarray}
\bar{P}&=&\sqrt{P}
\end{eqnarray}
The outer bound is a generalization of the AIS approach introduced in \cite{Arash_Jafar_GC14}. We will avoid repetition of  explanations for those steps that are essentially identical to \cite{Arash_Jafar_GC14}, and focus instead on the deviations from the original proof. 
As in \cite{Arash_Jafar_GC14}, the starting point  is to bound the problem with a deterministic model, such that a GDoF outer bound on the deterministic model is also a GDoF outer bound for the original problem. Since the derivation of the deterministic model is essentially identical to \cite{Arash_Jafar_GC14}, here we simply state the resulting deterministic model.

\subsubsection{Deterministic Channel Model}
{\bf Real Setting.}
$\forall k\in[K], t\in\mathbb{N}$, the deterministic channel model has inputs $\bar{X}_k(t) \in\mathbb{Z}$, and outputs $\bar{Y}_k(t) \in\mathbb{Z}$, such that
\begin{eqnarray}
\bar{Y}_k(t)&=&\sum_{l\in[K]/\{k\}}\lceil\bar{P}^{\alpha-\max(1,\alpha)}G_{kl}(t)\bar{X}_l(t)\rceil+\lceil \bar{P}^{1-\max(1,\alpha)}G_{kk}(t)\bar{X}_k(t)\rceil
\end{eqnarray}
such that
\begin{eqnarray}
\bar{X}_l(t)&\in&\{0,1,\cdots,\lceil \bar{P}^{\max(1,\alpha)} \rceil\}
\end{eqnarray}

\noindent{\bf Complex Setting.}
$\forall k\in[K], t\in\mathbb{N}$, the deterministic channel model has inputs $\bar{X}_k(t) \in\mathbb{C}$, and outputs $\bar{Y}_k(t) \in\mathbb{C}$, such that
\begin{eqnarray}
\bar{Y}_k(t)&=&\sum_{l\in[K]/\{k\}}\lceil\bar{P}^{\alpha-\max(1,\alpha)}G_{kl}(t)\bar{X}_l(t)\rceil+\lceil \bar{P}^{1-\max(1,\alpha)}G_{kk}(t)\bar{X}_k(t)\rceil
\end{eqnarray}
where the real and imaginary parts of the inputs, $\bar{X}_{kR}(t)$ and $\bar{X}_{kI}(t)$ are integers such that
\begin{eqnarray}
\bar{X}_{kR}(t),\bar{X}_{kI}(t)&\in&\{0,1,\cdots,\lceil \bar{P}^{\max(1,\alpha)} \rceil\}
\end{eqnarray}

%
\subsubsection{Key Lemmas}
The following lemmas are essential to the proof of the outer bound, and apply to both real and complex settings. 
\begin{lemma}\label{lemma1}
For $\alpha \leq 1$ we have
\begin{align}
&\forall k,l\in[K],~~H(\bar{Y}_{k}^{[n]}|G^{[n]})-H(G_{l,k}^{[n]}\bar{X}_k^{[n]}|G^{[n]})\leq  n\alpha C_o(P)+ o(\log{{P}})\label{lemma1-2}\\
&\forall k\in[K]/\{K\},~~ H(\bar{Y}_{k+1}^{[n]}|X_{1:k}^{[n]},G^{[n]})-H(\bar{Y}_k^{[n]}|X_{1:k}^{[n]},G^{[n]})\leq  n(1-\alpha)C_o(P)+ o(\log{{P}})\label{lemma1-0}
\intertext{For $\alpha>1$ we have,}
&\forall k\in[K]/\{K\},~~H(\bar{Y}_{k+1}^{[n]}|X_{1:k}^{[n]},G^{[n]})-H(\bar{Y}_k^{[n]}|X_{1:k}^{[n]},G^{[n]})\leq  o(\log{{P}})\label{lemma1-3}
\end{align}
\end{lemma}
The proof of Lemma \ref{lemma1} is presented in the Appendix. The following lemma follows directly from \cite{Arash_Jafar_GC14}.
\begin{lemma} \label{lemma2} Consider $\beta>0$, and random variables $F_k^n,G_k^n, k \in [K]$ that satisfy the bounded density assumption, i.e., they satisfy the conditions in Definition $(\ref{def1})$. Then, 
\begin{align}
& H\left(\sum_{k=1}^{K}\lceil{\bar{P}^{\beta}}F_{k}^{[n]}\bar{X}_k^{[n]}\rceil\right)\leq H\left(\sum_{k=1}^{K}\lceil{\bar{P}^{\beta}}G_{k}^{[n]}\bar{X}_k^{[n]}\rceil\right)+ o(\log{{P}})\label{lemma1-1}
\end{align}
\end{lemma}

\subsubsection{Proof of each of the 5 cases of Theorem \ref{Theorem1}}
Note that the following proof is applicable in both the real and complex settings. 
\allowdisplaybreaks
\begin{enumerate}
	\item{$\alpha\leq \frac{1}{2}, d(\alpha) \leq (1-\alpha)$.}\\
Starting from Fano's Inequality,
\begin{eqnarray}
n\sum_{k\in[K]}R_k&\leq&\sum_{k\in[K]} H(\bar{Y}_k^{[n]}|G^{[n]})-H(\bar{Y}_k^{[n]}|X_k^{[n]},G^{[n]})+n~o(n)\nonumber
\end{eqnarray}
Omitting for simplicity the $o(n)$ and $o(\log(P))$ terms which are inconsequential for GDoF calculations we proceed as follows,
\begin{align}
&=\sum_{k\in[K]} H\left(\sum_{l\in[K]/\{k\}}\lceil{\bar{P}^{\alpha-1}}G_{kl}^{[n]}\bar{X}_l^{[n]}\rceil+\lceil \left. G_{kk}^{[n]}\bar{X}_k^{[n]}\rceil\right|G^{[n]}\right)
-\sum_{k\in[K]}H\left(\sum_{l\in[K]/\{k\}}\lceil\left.{\bar{P}^{\alpha-1}}G_{kl}^{[n]}\bar{X}_l^{[n]}\rceil\right|G^{[n]}\right)\label{eq:Fano-1-1}\\
&\leq\sum_{k\in[K]} H\left(\sum_{l\in[K]/\{k\}}\lceil{\bar{P}^{\alpha-1}}G_{kl}^{[n]}\bar{X}_l^{[n]}\rceil+\lceil \left. G_{kk}^{[n]}\bar{X}_k^{[n]}\rceil\right|G^{[n]}\right)
-\sum_{k\in[K]}H\left(\lceil\left.{\bar{P}^{\alpha-1}}G_{k,k-1}^{[n]}\bar{X}_{k-1}^{[n]}\rceil\right|G^{[n]}\right)\label{eq:Fano-1-2}\\
&\leq\sum_{k\in[K]} H\left(\sum_{l\in[K]/\{k\}}\lceil{\bar{P}^{\alpha-1}}G_{kl}^{[n]}\bar{X}_l^{[n]}\rceil+\lceil \left. G_{kk}^{[n]}\bar{X}_k^{[n]}\rceil\right|G^{[n]}\right)
-\sum_{k\in[K]}H\left(\lceil\left.{\bar{P}^{\alpha-1}}G_{kk}^{[n]}\bar{X}_{k}^{[n]}\rceil\right|G^{[n]}\right)\label{eq:Fano-1-22}\\
&\leq\sum_{k\in[K]} H\left(\sum_{l\in[K]/\{k\}}\lceil{\bar{P}^{\alpha-1}}G_{kl}^{[n]}\bar{X}_l^{[n]}\rceil+\lceil \left. G_{kk}^{[n]}\bar{X}_k^{[n]}\rceil\right|\lceil{\bar{P}^{\alpha-1}}G_{kk}^{[n]}\bar{X}_{k}^{[n]}\rceil, G^{[n]}\right)\label{eq:Fano-1-3}\\
&\leq\sum_{k\in[K]} H\left(\sum_{l\in[K]/\{k\}}\lceil{\bar{P}^{\alpha-1}}G_{kl}^{[n]}\bar{X}_l^{[n]}\rceil+\lceil \left. G_{kk}^{[n]}\bar{X}_k^{[n]}\rceil-\lceil\bar{P}^{1-\alpha}\lceil{\bar{P}^{\alpha-1}}G_{kk}^{[n]}\bar{X}_{k}^{[n]}\rceil\rceil \right| G^{[n]}\right)\label{eq:Fano-1-4}\\
&\leq n~K~({1-\alpha})C_o(P)\label{eq:Fano-1-5}
\end{align}
where $(\ref{eq:Fano-1-1})$ is written by summing over all inequalities and  $(\ref{eq:Fano-1-2})$ came from independence of $X_i^n$ from $X_j^n$ for all $j\neq i$. $(\ref{eq:Fano-1-22})$ is true under finite precision CSIT due to Lemma {\ref{lemma2}}.  $(\ref{eq:Fano-1-3})$ comes from the fact that $H(A)-H(B)\leq H(A)+H(B|A)-H(B)=H(A,B)-H(B)=H(A|B)$. $(\ref{eq:Fano-1-4})$ is true because  for any function $f$, $H(A|B)=H(A-f(B)|B)\leq H(A-f(B))$. $(\ref{eq:Fano-1-5})$ is true because  $\lceil A\rceil - \lceil B \lceil A/B\rceil\rceil$ can only takes values between $0$ and $B+1$\footnote{$A/B\leq\lceil A/B\rceil < (A/B)+1$. Multiplying by $B$, we have $A\leq B\lceil A/B\rceil <A+B$, so that $\lceil A\rceil \leq \lceil B\lceil A/B\rceil\rceil < \lceil A\rceil + B+1.$}, so the range of values of the random variable within the parantheses is $O(\bar{P}^{1-\alpha})$, and finally the entropy of a random variable is bounded by the logarithm of its cardinality.

\item {$\frac{1}{2}<\alpha \leq \frac{K}{K+1}, d(\alpha) \leq\frac{1}{K-1}(K-2-(K-3)\alpha)$}\\
Starting with Fano's inequality, and omitting as before the $o(n)$ and $o(\log(P))$ terms,
\begin{align}
nR_k&\leq H(\bar{Y}_k^{[n]}|X_{[1:k-1]}^{[n]},G^{[n]})-H(\bar{Y}_k^{[n]}|X_{[1:k]}^{[n]},G^{[n]})\nonumber\\
n\sum_{k=1}^{K-1}R_k&\leq H(\bar{Y}_1^{[n]}|G^{[n]})+\sum_{k=1}^{K-2}\{H(\bar{Y}_{k+1}^{[n]}|X_{[1:k]}^{[n]},G^{[n]})-H(\bar{Y}_k^{[n]}|X_{[1:k]}^{[n]},G^{[n]})\}-H(\bar{Y}_{K-1}^{[n]}|X_{[1:K-1]}^{[n]},G^{[n]})\label{Fanoitem2-0}\\
&\leq H(\bar{Y}_1^{[n]}|G^{[n]})+n(K-2)(1-\alpha)C_o(P)-H(G_{K-1,K}^{[n]}\bar{X}_K^{[n]}) \label{case2-0}
\end{align}
where (\ref{case2-0}) came from Lemma \ref{lemma1}. Proceeding in the same manner, $K-1$ similar bounds can be written, for $1\leq m\leq K-1$, 
\begin{eqnarray}
n\sum_{k\in[K]/\{m\}}R_k&\leq&H(\bar{Y}_{m+1}^{[n]}|G^{[n]})+n(K-2)(1-\alpha)C_o(P)-H(G_{m-1,m}^{[n]}\bar{X}_m^{[n]}|G^{[n]})
\end{eqnarray}
Summing over all these $K$ bounds, we have,
\begin{align}
n\sum_{m\in[K]}\sum_{k\in[K]/\{m\}}R_k&\leq nK(K-2)(1-\alpha)C_o(P)+\sum_{m=1}^{K}\left\{H(\bar{Y}_{m+1}^{[n]}|G^{[n]})-H(G_{m-1,m}^{[n]}\bar{X}_m^{[n]}|G^{[n]})\right\}\\
&\leq nK(K-2)(1-\alpha)C_o(P)+\sum_{m=1}^{K}\left\{H(\bar{Y}_{m}^{[n]}|G^{[n]})-H(G_{m-1,m}^{[n]}\bar{X}_m^{[n]}|G^{[n]})\right\}\\
&\leq nK(K-2)(1-\alpha)C_o(P)+nK\alpha C_o(P)\label{case2-1}\\
&\leq nK\{K-2-(K-3)\alpha\}C_o(P)
\end{align}
where we interpret user indices modulo $K$, e.g., user $K+1$ is the same as user $1$. (\ref{case2-1}) is due to Lemma \ref{lemma1}. Normalizing throughout by $nC_o(P)$, in the GDoF limit, the left hand side  becomes $K(K-1)d(\alpha)$, producing the desired bound.

\item {$\frac{K}{K+1}<\alpha \leq 1, d(\alpha) \leq 1-\left(\frac{K-1}{K}\right)\alpha$}\\
In a similar fashion as (\ref{Fanoitem2-0}), and suppressing $o(n)$ and $o(\log(P))$ terms, we have,
\begin{eqnarray}
n\sum_{k=1}^{K} R_k&\leq&H(\bar{Y}_1^{[n]}|G^{[n]})+\sum_{k=1}^{K-1}\{H(\bar{Y}_{k+1}^{[n]}|X_{[1:k]}^{[n]},G^{[n]})-H(\bar{Y}_k^{[n]}|X_{[1:k]}^{[n]},G^{[n]})\}\\
&\leq&H(\bar{Y}_1^{[n]}|G^{[n]})+n(K-1)(1-\alpha)C_o(P)\label{case2-2}\\
&\leq&nC_o(P)+n(K-1)(1-\alpha)C_o(P)
\end{eqnarray}
Normalizing by $nC_o(P)$ and applying the GDoF limit, the bound on GDoF per user is obtained as $d(\alpha) \leq 1-\left(\frac{K-1}{K}\right)\alpha$.

\item {$1<\alpha \leq K, d(\alpha) \leq \alpha$}\\
In a similar fashion as (\ref{Fanoitem2-0}), and suppressing $o(n)$ and $o(\log(P))$ terms, we have,
\begin{eqnarray}
n\sum_{k=1}^{K} R_k&\leq&H(\bar{Y}_1^{[n]}|G^{[n]})+\sum_{k=1}^{K-1}\{H(\bar{Y}_{k+1}^{[n]}|X_{[1:k]}^{[n]},G^{[n]})-H(\bar{Y}_k^{[n]}|X_{[1:k]}^{[n]},G^{[n]})\}\\
&\leq&H(\bar{Y}_1^{[n]}|G^{[n]})\label{case2-3}\\
&\leq&n\alpha C_o(P)
\end{eqnarray}
where (\ref{case2-3}) came from Lemma \ref{lemma1}. Normalizing by $nC_o(P)$ and applying the GDoF limit, we have the outer bound on the GDoF per user as $d(\alpha)\leq \alpha$.

\item {$\alpha>K, d(\alpha)\leq 1$.} This bound is obvious from the  individual capacity bounds for each user in the absence of all interference.\\ 

\end{enumerate}

\subsection {Achievability}
Consider each of the cases of Theorem \ref{Theorem1}. We will present the proof for the real setting, but all the arguments apply directly to the complex setting as well.
\begin{enumerate}
\item{$\alpha\leq \frac{1}{2}, d(\alpha) \geq (1-\alpha)$.}\\
The $K$  messages are encoded into independent Gaussian codebooks $X_1,\cdots,X_K$ with powers E$|X_k|^2=P^{-\alpha}$. Suppressing the time index for clarity, the
received signals are:
\begin{eqnarray}
Y_k=\bar{P}G_{kk}X_k+\sum_{l\in[K]/\{k\}}\bar{P}^{\alpha}G_{kl}X_l+Z_k, \forall k \in [K]
\end{eqnarray}
Each receiver should decode its message successfully treating everything else as white noise.
The SINR value for this decoding is 
\begin{eqnarray}
\frac{PP^{-\alpha}|G_{kk}|^2}{1+\sum_{l\in[K]/\{k\}}P^{\alpha}P^{-\alpha}|G_{kl}|^2}\geq \frac{P^{1-\alpha}\Delta_1^2}{1+(K-1)\Delta_2^2}
\end{eqnarray}
so that the achievable rate (for real channels) per user is $0.5\log(1+$SINR$) = 0.5(1-\alpha)\log(P) $+$~o(\log(P)) = (1-\alpha) \log( \bar{P})$ +$~o(\log(P))$ which shows that  $(1-\alpha)$ GDoF per user are achievable, simply by treating interference as noise.\\

\item{$ \frac{1}{2}<\alpha\leq \frac{K}{K+1}, d(\alpha) \geq \frac{1}{K-1}(K-2-(K-3)\alpha)$.}\\
Each message is divided into two sub-messages, a private message and a public message. Each private message is intended to carry $1-\alpha$ GDoF while each public message is intended to carry $\frac{2\alpha-1}{K-1}$ GDoF, for a total of $1-\alpha+\frac{2\alpha-1}{K-1}=\frac{1}{K-1}(K-2-(K-3)\alpha)$ GDoF per user.

The private messages are encoded into Gaussian codebooks $V_1,\cdots,V_K$ with powers E$|V_k|^2=P^{-\alpha}$. The public messages are encoded into Gaussian codebooks $U_1,\cdots,U_K$ with powers E$|U_k|^2=1-P^{-\alpha}$ so that the total power per transmitter is unity. Suppressing the time index for simplicity, the
received signals are:
\begin{eqnarray}
Y_k=\bar{P}G_{kk}(V_k+U_k)+\sum_{l\in[K]/\{k\}}\bar{P}^{\alpha}G_{kl}(V_l+W_l)+Z_k, \forall k \in [K]
\end{eqnarray}
Without loss of generality, consider Receiver 1. It first decodes $U_1$ treating everything else as noise.
The SINR value for this decoding is 
\begin{eqnarray}
\frac{P(1-P^{-\alpha})|G_{11}|^2}{1+PP^{-\alpha}|G_{11}|^2+\sum_{l=2}^{K}P^{\alpha}(1-P^{-\alpha})|G_{1l}|^2+\sum_{l=2}^{K}P^{\alpha}P^{-\alpha}|G_{1l}|^2}\geq \frac{P^{1-\alpha}\Delta_1^2}{(K-1)\Delta_2^2}
\end{eqnarray}
Since $\frac{2\alpha-1}{K-1}\leq 1-\alpha$, Receiver 1 decodes $U_1$ successfully and subtracts it from its received signal to obtain a new received signal $Y_1'$. In the next step it acts as a multiple access receiver, and \emph{jointly} decodes all remaining $U_k, \forall k \in \{2,3,\cdots,K\}$, while treating all $V_k$ as white noise.
\begin{eqnarray}
Y_1'=\sum_{k=2}^{K}\bar{P}^{\alpha}G_{1k}U_k+N_1\\
N_1=\bar{P}G_{11}V_1+\sum_{k=2}^{K}\bar{P}^{\alpha}G_{1k}V_k+Z_1 \label{ac}
\end {eqnarray}
This is a Gaussian MAC with $K-1$ messages where each desired signal has power $\sim P^{\alpha}$ while the Gaussian noise has power $\sim P^{1-\alpha}$. Thus, all $K-1$ messages, each carrying $\frac{2\alpha-1}{K-1}$ are decoded successfully.
After decoding all public messages $U_k$ and subtracting them from its received signal, Receiver 1 decodes its private message $V_1$ by treating all other $V_j$ as noise. The SINR for this decoding is,
\begin{eqnarray}
&&\frac{P(P^{-\alpha})|G_{11}|^2}{1+\sum_{j=1,j\neq 1}^{K}P^{\alpha}P^{-\alpha}|G_{1j}|^2}\\
&\geq& \frac{P^{1-\alpha}\Delta_1^2}{(K-1)\Delta_2^2}
\end {eqnarray}
Therefore, the private message which carries $1-\alpha$ GDoF is also decoded successfully.

\item {$\frac{K}{K+1}<\alpha \leq 1, d(\alpha) \leq 1-\left(\frac{K-1}{K}\right)\alpha$}\\
Each message is divided into  a private message and a public message. Each private message is intended to carry $1-\alpha$ GDoF while each public message is intended to carry $\alpha/K$ GDoF, for a total of $1-\alpha+\frac{\alpha}{K}=1-\left(\frac{K-1}{K}\right)\alpha$ GDoF per user. The private messages are encoded into Gaussian codebooks $V_1,\cdots,V_K$ with powers E$|V_k|^2=P^{-\alpha}$. The public messages are encoded into Gaussian codebooks $U_1,\cdots,U_K$ with powers E$|U_k|^2=1-P^{-\alpha}$ so that the total power per transmitter is unity. 

Without loss of generality, consider Receiver 1. It first jointly decodes all public messages $U_1, \cdots, U_K$, while treating all private messages as white noise. The GDoF region for this multiple access channel is the following.
\begin{eqnarray}
\{(d_{U_1},d_{U_2}, \cdots, d_{U_K}):&& \forall \mathcal{M}\subset \{2,3,\cdots, K\},\\
&&\sum_{m\in\mathcal{M}}d_{U_m}\leq \alpha - (1-\alpha),\\
&&\sum_{m\in\mathcal{M}\cup\{1\}}d_{U_m}\leq 1 - (1-\alpha)\}.
\end{eqnarray}
Since $\forall k\in[K]$, $d_{U_k}=\alpha/K$ belongs to the GDoF region of the multiple access channel, the public messages are successfully decoded. After subtracting the contributions of the public messages from its received signal, Receiver 1 is able to decode its private message by treating all other remaining signals as  white noise, similar to the case $\frac{1}{2}<\alpha\leq\frac{K}{K+1}$.

\item {$1<\alpha, d(\alpha) \leq \min(\frac{\alpha}{K},1)$}\\
All $K$ messages are public in this case and are encoded into Gaussian codebooks $U_1,\cdots,U_K$ with powers E$|U_i|^2=1$, each carrying $\min(\frac{\alpha}{K},1)$ GDoF. Without loss of generality, consider Receiver 1, which sees the signal,
\begin{eqnarray}
Y_1=\bar{P}G_{11}U_1+\sum_{k=2}^{K}\bar{P}^{\alpha}G_{1k}U_k+Z_1
\end {eqnarray}
The GDoF region for this multiple access channel is the following.
\begin{eqnarray}
\{(d_{U_1},d_{U_2}, \cdots, d_{U_K}):&&d_{U_1}\leq 1 \\
&& \forall \mathcal{M}\subset [K],\sum_{m\in\mathcal{M}}d_{U_m}\leq \alpha.\}.
\end{eqnarray}
Since $\forall k\in[K]$, $d_{U_k}=\min(\frac{\alpha}{K},1)$ belongs to the GDoF region of the multiple access channel, Receiver 1 is able to successfully decode all the messages.
\end{enumerate}

\section{Conclusion}
We characterized the GDoF of the  K-user symmetric interference channel under finite precision CSIT. Unlike asymmetric settings studied under the TIM framework, in the symmetric setting the benefits of interference alignment are lost. To achieve the optimal GDoF, it suffices to partition each message into a private and a public component, so that the private component is treated as white noise at all undesired receivers, while the public messages are jointly decoded at each receiver. A highlight of this work is the extension of the Aligned Image Sets (AIS) approach to obtain tight GDoF outer bounds for the symmetric $K$ user setting. Extensions to asymmetric settings remain an intriguing direction for future work.

\section{Appendix -- Proof of Lemma \ref{lemma1}}
\subsection{Proof of Bound (\ref{lemma1-2})}
Let us prove $(\ref{lemma1-2})$ for both  real and complex settings. Without loss of generality, suppose $k=1$.
\begin {align}
\intertext{$H(\bar{Y}_{1}^{[n]}|G^{[n]})-H(G_{K,1}^{[n]}\bar{X}_1^{[n]}|G^{[n]})$}
&\leq H(\bar{Y}_{1}^{[n]}|G^{[n]})-H(G_{1,1}^{[n]}\bar{X}_1^{[n]}|G^{[n]})+~o(\log{{P}})\label{pl1}\\
&\leq H(\bar{Y}_{1}^{[n]}|G_{1,1}^{[n]}\bar{X}_1^{[n]},G^{[n]})+~o(\log{{P}})\label{pl2}\\
&= H\left(\left.\sum_{k=2}^{K}\lceil{\bar{P}^{\alpha-1}}G_{1k}^{[n]}\bar{X}_k^{[n]}\rceil+\lceil G_{11}^{[n]}\bar{X}_1^{[n]}\rceil~\right|~G_{1,1}^{[n]}\bar{X}_1^{[n]},G^{[n]}\right)+~o(\log{{P}})\label{pl3}\\
&= H\left(\left.\sum_{k=2}^{K}\lceil{\bar{P}^{\alpha-1}}G_{1k}^{[n]}\bar{X}_k^{[n]}\rceil\right|G^{[n]}\right)+~o(\log{{P}})\label{pl4}\\
&\leq  n\alpha C_o(P)+ ~o(\log{{P}})\label{pl5}
\end{align}
where $(\ref{pl1})$ is due to Lemma \ref{lemma2}). $(\ref{pl3})$ and $(\ref{pl4})$ are  just rewriting $\bar{Y}_{1}^{[n]}$ and using the fact that removing the conditioning does not decrease the entropy. $(\ref{pl5})$ is  true because the  entropy of a random variable can be bounded by logarithm of its cardinality.\\

\subsection{Proof of Bound (\ref{lemma1-0})}
\subsubsection{Real Setting}
Now let us prove  $(\ref{lemma1-0})$, first for the real setting.  Without loss of generality, let $k=1$, so we need to prove the following.
\begin {eqnarray}
H(\bar{Y}_{2}^{[n]}|X_1^{[n]},G^{[n]})-H(\bar{Y}_1^{[n]}|X_1^{[n]},G^{[n]})&\leq&  n(1-\alpha)\log{\bar{P}}+ o~(\log{{P}})\label{eq:lemma1-0new}
\end {eqnarray}

\noindent{\bf  Functional Dependence.}
As in \cite{Arash_Jafar_GC14}, for a given channel realization for user $2$, $G_2^{[n]}$, and a given $\bar{X}_1^{[n]}$, there are multiple vectors $(\bar{X}_2^{[n]}, \bar{X}_3^{[n]},\cdots, \bar{X}_K^{[n]})$ that cast the same image in $\bar{Y}_2^{[n]}$. Thus, given $\bar{X}_1^{[n]}$ and the channel for user $2$, the mapping, $\mathcal{L}$, from $\bar{Y}_2^{[n]}$ to one of these vectors $(\bar{X}_2^{[n]}, \bar{X}_3^{[n]},\cdots, \bar{X}_K^{[n]})$ is a random variable. Conditioning on this mapping $H(\bar{Y}_1^{[n]}|X_1^{[n]},G^{[n]})\geq H(\bar{Y}_1^{[n]}|X_1^{[n]},G^{[n]},\mathcal{L})$, and then instead of averaging over $\mathcal{L}$, choosing the mapping $\mathcal{L}_o$ that minimizes this entropy term provides an outer bound on $H(\bar{Y}_{2}^{[n]}|X_1^{[n]},G^{[n]})-H(\bar{Y}_1^{[n]}|X_1^{[n]},G^{[n]})$, with the condition that now
\begin{eqnarray}
(\bar{X}_{2}^{[n]},\cdots, \bar{X}_K^{[n]})&=&\mathcal{L}_o(\bar{Y}_{2}^{[n]},\bar{X}_1^{[n]}, G_2^{[n]})\\
\Rightarrow \bar{Y}_1^{[n]}&=&f(\bar{Y}_{2}^{[n]},\bar{X}_1^{[n]}, G^{[n]})
\end{eqnarray}
where $a=f(b)$ denotes that $a$ is \emph{some} function of $b$.\\

\noindent{\bf Aligned Image Sets.} Given $\bar{X}_1^{[n]}$ and channel realizations $G^{[n]}$, define $S_{\bar{Y_2}^{[n]}}(G^{[n]},\bar{X}_1^{[n]})$ as the set of all codewords $(\bar{X}_2^{[n]},\cdots,\bar{X}_K^{[n]})$ that produce the same output, $\bar{Y_1}^{[n]}$,  at Receiver $1$, as is produced at Receiver $1$ by the codeword $\mathcal{L}_o(\bar{Y}_{2}^{[n]},\bar{X}_1^{[n]}, G_2^{[n]})$.
\begin{align}
H(\bar{Y_2}^{[n]}, S_{\bar{Y_2}^{[n]}}|\bar{X}_1^{[n]},G^{[n]})&=H(\bar{Y_2}^{[n]}|\bar{X}_1^{[n]},G^n)+H(S_{\bar{Y_2}^{[n]}}|\bar{X}_1^{[n]},G^{[n]}, \bar{Y_2}^{[n]})\\
&=H(\bar{Y_2}^{[n]}|\bar{X}_1^{[n]},G^n)\label{eq:aligned1}\\
&=H(S_{\bar{Y_2}^{[n]}}|\bar{X}_1^{[n]},G^{[n]})+H(\bar{Y_2}^{[n]}|S_{\bar{Y_2}^{[n]}},\bar{X}_1^{[n]},G^{[n]})\nonumber\\
&=H({\bar{Y_1}^{[n]}}|\bar{X}_1^{[n]},G^{[n]})+H(\bar{Y_2}^{[n]}|S_{\bar{Y_2}^{[n]}},\bar{X}_1^{[n]},G^{[n]})\\
&\leq H({\bar{Y_1}^{[n]}}|\bar{X}_1^{[n]},G^{[n]})+\mbox{E}[\log|S_{\bar{Y_2}^{[n]}}|]\\
&\leq H({\bar{Y_1}^{[n]}}|\bar{X}_1^{[n]},G^{[n]})+\log\left(\mbox{E}[|S_{\bar{Y_2}^{[n]}}|]\right)\label{eq:aligned2}
\end{align}
From (\ref{eq:aligned1}) and (\ref{eq:aligned2}) we have
\begin{eqnarray}
H(\bar{Y_2}^{[n]}|W,G^n)-H({\bar{Y_1}^{[n]}}|W,G^{[n]})&\leq&\log\left(\mbox{E}[|S_{\bar{Y_2}^{[n]}}|]\right)\label{ss}
\end{eqnarray}
So it only remains to bound the average size of an aligned image set, $\mbox{E}[|S_{\bar{Y_2}^{[n]}}|]$. Proceeding as in \cite{Arash_Jafar_GC14},
\begin{eqnarray}
\mbox{E}[|S_{\bar{y}_2^{[n]}}|]&=&\sum_{{\bar{y'}_2}^n\in\{\bar{Y_2}^{[n]}\}}\mathbb{P}\left({\bar{y'}_2}^n\in S_{\bar{y}_2^{[n]}}\right)
\end{eqnarray}
\noindent{\bf Probability that Images Align.}
Given $G_{21}^{[n]},\cdots, G_{2K}^{[n]}$, consider two distinct realizations of Receiver 2's output sequence $\bar{Y}_2^{[n]}$, denoted as $\lambda^{[n]}$ and $\nu^{[n]}$, which are produced by the corresponding two realizations of the codeword $(X_2^{[n]},\cdots, X_K^{[n]})$ denoted by $(\lambda_2^{[n]},\cdots,\lambda_K^{[n]})$ and $(\nu_2^{[n]},\cdots, \nu_K^{[n]})$, respectively. 
\begin{eqnarray}
\lambda(t)&=&\lfloor G_{22}(t)\lambda_2(t)\rfloor+\sum_{k=3}^{K}\lfloor \bar{P}^{\alpha-1}G_{2k}(t)\lambda_k(t)\rfloor\label{eq:lambdanu1}\\
\nu(t)&=&\lfloor G_{22}(t)\nu_2(t)\rfloor+\sum_{k=3}^{K}\lfloor \bar{P}^{\alpha-1}G_{2k}(t)\nu_k(t)\rfloor\label{eq:lambdanu2}
\end{eqnarray}
We wish to bound the probability that the images of these two codewords align at Receiver 1, i.e., $\lambda^{[n]}\in S_{\nu^{[n]}}$. For simplicity, consider first the single channel use setting, $n=1$. For $\lambda\in S_\nu$ we must have
\begin{eqnarray}
\sum_{j=2}^{K}\lfloor \bar{P}^{\alpha-1}G_{1j}(t)\lambda_j(t)\rfloor
=\sum_{j=2}^{K}\lfloor \bar{P}^{\alpha-1}G_{1j}(t)\nu_j(t)\rfloor
\end{eqnarray}
Proceeding as in \cite{Arash_Jafar_GC14}, the probability of $\nu\in S_\lambda$ is no more than $\frac{2\bar{P}^{1-\alpha}f_{\max}}{\max_i|\lambda_i-\nu_i|}$, so that, 
\begin{align}
\mathbb{P}(\lambda\in S_\nu)&\leq 
\left\{
\begin{array}{ll}
\frac{2 (K-1)\Delta_2\bar{P}^{1-\alpha}f_{\max}}{|\lambda-\nu|-(K-1)},&\mbox{if }|\lambda-\nu|>(K-1)
\intertext{\color{red} }
1,&\mbox{otherwise.}
\end{array}
\right.
\end{align}
Now let us return to the case of general $n$, where we similarly have,
\begin{eqnarray}
\mathbb{P}(\lambda^{[n]}\in S_{\nu^{[n]}})&\leq&\max(2(K-1)\Delta_2f_{\max},1)^n\bar{P}^{n(1-\alpha)}\nonumber\\
&&\times \prod_{t:|\lambda(t)-\nu(t)|>K-1}\frac{1}{|\lambda(t)-\nu(t)|-K+1}
\end{eqnarray}
{\bf Bounding the Expected Size of Aligned Image Sets.}
\begin{eqnarray}
\mbox{E}(|S_{\nu^{[n]}}|)&=&\sum_{\lambda^n\in\{\bar{Y_2}^{[n]}\}}\mathbb{P}\left(\lambda^n\in S_{\nu^{[n]}}\right)\\
&\leq&\max(2(K-1)\Delta_2f_{\max},1)^n\bar{P}^{n(1-\alpha)}\nonumber\\
&&\prod_{t=1}^n\left(\sum_{\lambda(t): |\lambda(t)-\nu(t)|\leq K-1}1+\sum_{\lambda(t): K-1< |\lambda(t)-\nu(t)|\leq Q_y}\frac{1}{|\lambda(t)-\nu(t)|-K+1}\right)\nonumber\\
&\leq&\max(2(K-1)\Delta_2f_{\max},1)^n\bar{P}^{n(1-\alpha)}\times\left(\log(\bar{P})+o(\log(\bar{P}))\right)^n\label{eq:aveSK}
\end{eqnarray}
where $Q_y\leq(2\Delta_2P+1)$. 
Substituting $(\ref{eq:aveSK})$ back into (\ref{ss}), $(\ref{lemma1-0})$ is obtained.

\subsubsection{Complex Setting}
The proof is for the most part straightforward based on the real setting studied earlier. To avoid repetition, here we highlight only the differences.

\noindent{\bf  Probability that Images Align.}
Given $G_{21}^{[n]},\cdots, G_{2K}^{[n]}$, consider two distinct realizations of user 2's output sequence $\bar{Y}_2^{[n]}$, denoted as $\lambda^{[n]}$ and $\nu^{[n]}$, which are produced by the corresponding two realizations of the codeword $(X_2^{[n]},\cdots, X_K^{[n]})$ denoted by $(\lambda_2^{[n]},\cdots,\lambda_K^{[n]})$ and $(\nu_2^{[n]},\cdots, \nu_K^{[n]})$, respectively. 
\begin{eqnarray}
\lambda(t)&=&\lfloor (G_{22,R}(t)+jG_{22,I}(t))(\lambda_{2,R}(t)+j\lambda_{2,I}(t))\rfloor+\nonumber\\
&&\sum_{k=3}^{K}\lfloor \bar{P}^{\alpha-1}(G_{2k,R}(t)+jG_{2k,I}(t))(\lambda_{k,R}(t)+j\lambda_{k,I}(t))\rfloor\label{eq:lambdanu1c}\\
\nu(t)&=&\lfloor (G_{22,R}(t)+jG_{22,I}(t))(\nu_{2,R}(t)+j\nu_{2,I}(t))\rfloor\nonumber\\
&&+\sum_{k=3}^{K}\lfloor \bar{P}^{\alpha-1}(G_{2k,R}(t)+jG_{2k,I}(t))(\nu_{k,R}(t)+j\nu_{k,I}(t))\rfloor\label{eq:lambdanu2c}
\end{eqnarray}
We wish to bound the probability that the images of these two codewords align at User 1, i.e., $\lambda^{[n]}\in S_{\nu^{[n]}}$. For simplicity, consider first a single channel use, $n=1$. For $\lambda\in S_\nu$ we must have
\begin{eqnarray}
&&\sum_{k=2}^{K}\lfloor \bar{P}^{\alpha-1}(G_{1k,R}(t)+jG_{1k,I}(t))(\lambda_{k,R}(t)+j\lambda_{k,I}(t))\rfloor\nonumber\\
&=&\sum_{k=2}^{K}\lfloor \bar{P}^{\alpha-1}(G_{1k,R}(t)+jG_{1k,I}(t))(\nu_{k,R}(t)+j\nu_{k,I}(t))\rfloor
\end{eqnarray}
Proceeding as in \cite{Arash_Jafar_GC14},  the probability of $\lambda\in S_\nu$ is no more than $(\frac{16f_{\max}^2P^{1-\alpha}}{(\max_i\max(|\lambda_{i,R}-\nu_{i,R}|,|\lambda_{i,I}-\nu_{i,I}|))^2})$. So, for $\max(\Re{|\lambda-\nu|},\Im{|\lambda-\nu|})> K-1$ the probability of alignment is no more than

\begin{eqnarray}
\frac{64(K-1)^2\Delta_2^2f_{\max}^2{P}^{1-\alpha}}{\max(\Re{|\lambda-\nu|}-K+1,1)\max(\Im{|\lambda-\nu|}-K+1,1)}
\end{eqnarray}
For $\max(\Re{|\lambda-\nu|},\Re{|\lambda-\nu|}) \leq K-1$ the probability of alignment is bounded by unity. Now let us return to the case of general $n$, where we similarly have,
\begin{eqnarray}
\mathbb{P}(\lambda^{[n]}\in S_{\nu^{[n]}})&\leq&\max(1,64(K-1)^2\Delta_2^2f_{\max}^2)^n{P}^{n(1-\alpha)}\times \prod_{t:\max (\Re {|\lambda(t)-\nu(t)|},\Im {|\lambda(t)-\nu(t)|})>K-1}\nonumber\\
&&\frac{1}{\max(\Re {|\lambda(t)-\nu(t)|}-K+1,1)\max(\Im {|\lambda(t)-\nu(t)|}-K+1,1)}
\end{eqnarray}
{\bf  Bounding the Expected Size of Aligned Image Sets.}
\begin{eqnarray}
\mbox{E}(|S_{\nu^{[n]}}|)&=&\sum_{\lambda^n\in\{\bar{Y_1}^{[n]}\}}\mathbb{P}\left(\lambda^n\in S_{\nu^{[n]}}\right)\\
&\leq&\max(1,64(K-1)^2\Delta_2^2f_{\max}^2)^n{P}^{n(1-\alpha)}\nonumber\\
&&\times\prod_{t=1}^n\left(\sum_{\Re {|\lambda(t)-\nu(t)|}: \Re {|\lambda(t)-\nu(t)|}\leq K-1}1+\sum_{\Re {|\lambda(t)-\nu(t)|}: K-1< \Re {|\lambda(t)-\nu(t)|}\leq Q_y}\frac{1}{\Re {|\lambda(t)-\nu(t)|}-K+1}\right)\nonumber\\
&&\times\prod_{t=1}^n\left(\sum_{\Im {|\lambda(t)-\nu(t)|}: \Im {|\lambda(t)-\nu(t)|}\leq K-1}1+\sum_{\Im {|\lambda(t)-\nu(t)|}: K-1< \Im {|\lambda(t)-\nu(t)|}\leq Q_y}\frac{1}{\Im {|\lambda(t)-\nu(t)|}-K+1}\right)\nonumber\\
&\leq&\max(1,64(K-1)^2\Delta_2^2f_{\max}^2)^n{P}^{n(1-\alpha)}\times\left(C_o(P)+o(\log({{P}}))\right)^{2n}\nonumber\label{eq:avee}
\end{eqnarray}
where $Q_y\leq(3\Delta_2+K-1)\lceil  {\bar{P}}\rceil$. 
Substituting $(\ref{eq:avee})$ back into (\ref{ss}), $(\ref{lemma1-0})$ is obtained.\\

\subsection{Proof of Bound (\ref{lemma1-3})}
\subsubsection{Real and Complex Setting}
Without loss of generality, let $k=1$, so we need to prove the following.
\begin {eqnarray}
H(\bar{Y}_{2}^{[n]}|X_1^{[n]},G^{[n]})-H(\bar{Y}_1^{[n]}|X_1^{[n]},G^{[n]})&\leq&  o~(\log{{P}})\label{eq:lemma1-3new}
\end {eqnarray}

Functional Dependence and Aligned Image Sets are defined similar to the Proof of Bound (\ref{lemma1-0}).\\

\noindent{\bf Probability that Images Align.}
Given $G_{21}^{[n]},\cdots, G_{2K}^{[n]}$, consider two distinct realizations of Receiver 2's output sequence $\bar{Y}_2^{[n]}$, denoted as $\lambda^{[n]}$ and $\nu^{[n]}$, which are produced by the corresponding two realizations of the codeword $(X_2^{[n]},\cdots, X_K^{[n]})$ denoted by $(\lambda_2^{[n]},\cdots,\lambda_K^{[n]})$ and $(\nu_2^{[n]},\cdots, \nu_K^{[n]})$, respectively. 
\begin{eqnarray}
\lambda(t)&=&\lfloor \bar{P}^{1-\alpha} G_{22}(t)\lambda_2(t)\rfloor+\sum_{k=3}^{K}\lfloor G_{2k}(t)\lambda_k(t)\rfloor\label{eq:lambdanu1new}\\
\nu(t)&=&\lfloor \bar{P}^{1-\alpha} G_{22}(t)\nu_2(t)\rfloor+\sum_{k=3}^{K}\lfloor G_{2k}(t)\nu_k(t)\rfloor\label{eq:lambdanu2new}
\end{eqnarray}
We wish to bound the probability that the images of these two codewords align at Receiver 1, i.e., $\lambda^{[n]}\in S_{\nu^{[n]}}$. For simplicity, consider first the single channel use setting, $n=1$. For $\lambda\in S_\nu$ we must have
\begin{eqnarray}
\sum_{j=2}^{K}\lfloor G_{1j}(t)\lambda_j(t)\rfloor
=\sum_{j=2}^{K}\lfloor G_{1j}(t)\nu_j(t)\rfloor
\end{eqnarray}
Proceeding as in \cite{Arash_Jafar_GC14}, the probability of $\nu\in S_\lambda$ is no more than $\frac{2f_{\max}}{\max_i|\lambda_i-\nu_i|}$, so that, 
\begin{align}
\mathbb{P}(\lambda\in S_\nu)&\leq 
\left\{
\begin{array}{ll}
\frac{2 (K-1)\Delta_2f_{\max}}{|\lambda-\nu|-(K-1)},&\mbox{if }|\lambda-\nu|>(K-1)
\intertext{\color{red} }
1,&\mbox{otherwise.}
\end{array}
\right.
\end{align}
Now let us return to the case of general $n$, where we similarly have,
\begin{eqnarray}
\mathbb{P}(\lambda^{[n]}\in S_{\nu^{[n]}})&\leq&\max(2(K-1)\Delta_2f_{\max},1)^n\nonumber\\
&&\times \prod_{t:|\lambda(t)-\nu(t)|>K-1}\frac{1}{|\lambda(t)-\nu(t)|-K+1}
\end{eqnarray}
{\bf Bounding the Expected Size of Aligned Image Sets.}
\begin{eqnarray}
\mbox{E}(|S_{\nu^{[n]}}|)&=&\sum_{\lambda^n\in\{\bar{Y_2}^{[n]}\}}\mathbb{P}\left(\lambda^n\in S_{\nu^{[n]}}\right)\\
&\leq&\max(2(K-1)\Delta_2f_{\max},1)^n\nonumber\\
&&\prod_{t=1}^n\left(\sum_{\lambda(t): |\lambda(t)-\nu(t)|\leq K-1}1+\sum_{\lambda(t): K-1< |\lambda(t)-\nu(t)|\leq Q_y}\frac{1}{|\lambda(t)-\nu(t)|-K+1}\right)\nonumber\\
&\leq&\max(2(K-1)\Delta_2f_{\max},1)^n\times\left(\log(\bar{P})+o(\log(\bar{P}))\right)^n\label{eq:aveSK3}
\end{eqnarray}
where $Q_y\leq(2\Delta_2P+1)$. 
Substituting $(\ref{eq:aveSK3})$ back into (\ref{ss}), $(\ref{lemma1-3})$ is obtained.

\bibliographystyle{IEEEtran}
\bibliography{Thesis}

\begin{thebibliography}{10}
\providecommand{\url}[1]{#1}
\csname url@samestyle\endcsname
\providecommand{\newblock}{\relax}
\providecommand{\bibinfo}[2]{#2}
\providecommand{\BIBentrySTDinterwordspacing}{\spaceskip=0pt\relax}
\providecommand{\BIBentryALTinterwordstretchfactor}{4}
\providecommand{\BIBentryALTinterwordspacing}{\spaceskip=\fontdimen2\font plus
\BIBentryALTinterwordstretchfactor\fontdimen3\font minus
  \fontdimen4\font\relax}
\providecommand{\BIBforeignlanguage}[2]{{%
\expandafter\ifx\csname l@#1\endcsname\relax
\typeout{** WARNING: IEEEtran.bst: No hyphenation pattern has been}%
\typeout{** loaded for the language `#1'. Using the pattern for}%
\typeout{** the default language instead.}%
\else
\language=\csname l@#1\endcsname
\fi
#2}}
\providecommand{\BIBdecl}{\relax}
\BIBdecl

\bibitem{ADT_FnT}
A.~Avestimehr, S.~Diggavi, C.~Tian, and D.~Tse, ``An approximation approach to
  network information theory,'' in \emph{Foundations and Trends in
  Communication and Information Theory}, vol.~12, 2015, pp. 1--183.

\bibitem{Jafar_FnT}
S.~Jafar, ``Interference alignment: A new look at signal dimensions in a
  communication network,'' in \emph{Foundations and Trends in Communication and
  Information Theory}, vol.~7, 2011, pp. 1--136.

\bibitem{Cadambe_Jafar_int}
V.~Cadambe and S.~Jafar, ``Interference alignment and the degrees of freedom of
  the {K} user interference channel,'' \emph{IEEE Transactions on Information
  Theory}, vol.~54, no.~8, pp. 3425--3441, Aug. 2008.

\bibitem{Nazer_Gastpar_Jafar_Vishwanath}
B.~Nazer, M.~Gastpar, S.~Jafar, and S.~Vishwanath, ``Ergodic interference
  alignment,'' \emph{IEEE Trans. on Information Theory}, vol.~58, no.~10, pp.
  6355--6371, Oct. 2012.

\bibitem{Motahari_Gharan_Khandani}
A.~Motahari, S.~Gharan, M.~Maddah-Ali, and A.~Khandani, ``Real interference
  alignment: Exploiting the potential of single antenna systems,'' \emph{IEEE
  Transactions on Information Theory}, vol.~60, no.~8, pp. 4799 -- 4810, 2014.

\bibitem{Wu_Shamai_Verdu}
Y.~Wu, S.~Shamai, and S.~Verdu, ``Information dimension and the degrees of
  freedom of the interference channel,'' \emph{IEEE Transactions on Information
  Theory}, vol.~61, no.~1, pp. 256--279, Jan. 2015.

\bibitem{tse_mobicom}
D.~Tse, ``Breaking the interference barrier,'' in \emph{Mobicom/Mobihoc
  Plenary}, Sep 2007.

\bibitem{Lapidoth_Shamai_Wigger_BC}
A.~Lapidoth, S.~Shamai, and M.~Wigger, ``On the capacity of fading {MIMO}
  broadcast channels with imperfect transmitter side-information,'' in
  \emph{Proceedings of 43rd Annual Allerton Conference on Communications,
  Control and Computing}, Sep. 28-30, 2005.

\bibitem{Arash_Jafar_GC14}
A.~G. Davoodi and S.~Jafar, ``Settling conjectures on the collapse of degrees
  of freedom under finite precision {CSIT},'' \emph{Globecom (full paper at
  arXiv:1403.1541)}, Dec. 2014.

\bibitem{Etkin_Tse_Wang}
R.~Etkin, D.~Tse, and H.~Wang, ``{Gaussian interference channel capacity to
  within one bit},'' \emph{IEEE Transactions on Information Theory}, vol.~54,
  no.~12, pp. 5534--5562, 2008.

\bibitem{Jafar_scalar}
S.~Jafar and A.~Goldsmith, ``Isotropic fading vector broadcast channels: the
  scalar upperbound and loss in degrees of freedom,'' \emph{IEEE Trans. Inform.
  Theory}, vol.~51, no.~3, pp. 848--857, March 2005.

\bibitem{Huang_Jafar_Shamai_Vishwanath}
C.~Huang, S.~A. Jafar, S.~Shamai, and S.~Vishwanath, ``{On Degrees of Freedom
  Region of MIMO Networks without Channel State Information at Transmitters},''
  \emph{IEEE Transactions on Information Theory}, no.~2, pp. 849--857, Feb.
  2012.

\bibitem{Zhu_Guo_MIMOIC}
Y.~Zhu and D.~Guo, ``The degrees of freedom of isotropic {MIMO} interference
  channels without state information at the transmitters,'' \emph{IEEE
  Transactions on Information Theory}, vol.~58, no.~1, pp. 341--352, 2012.

\bibitem{Varanasi_noCSIT}
\BIBentryALTinterwordspacing
C.~S. Vaze and M.~K. Varanasi, ``The degrees of freedom regions of {MIMO}
  broadcast, interference, and cognitive radio channels with no {CSIT},''
  \emph{CoRR}, vol. abs/0909.5424, 2009. [Online]. Available:
  \url{http://arxiv.org/abs/0909.5424}
\BIBentrySTDinterwordspacing

\bibitem{Weingarten_Shamai_Kramer}
H.~Weingarten, S.~Shamai, and G.~Kramer, ``On the compound {MIMO} broadcast
  channel,'' in \emph{Proceedings of Annual Information Theory and Applications
  Workshop UCSD}, Jan 2007.

\bibitem{Gou_Jafar_Wang}
T.~Gou, S.~Jafar, and C.~Wang, ``On the degrees of freedom of finite state
  compound wireless networks,'' \emph{IEEE Transactions on Information Theory},
  vol.~57, no.~6, pp. 3268--3308, June 2011.

\bibitem{Maddah_Compound}
\BIBentryALTinterwordspacing
M.~A. Maddah-Ali, ``The degrees of freedom of the compound {MIMO} broadcast
  channels with finite states,'' \emph{CoRR}, vol. abs/0909.5006, 2009.
  [Online]. Available: \url{http://arxiv.org/abs/0909.5006}
\BIBentrySTDinterwordspacing

\bibitem{Tandon_Jafar_Shamai_Poor}
R.~Tandon, S.~A. Jafar, S.~Shamai, and H.~V. Poor, ``On the synergistic
  benefits of alternating \textsc{CSIT} for the \textsc{MISO} \textsc{BC},''
  \emph{IEEE Transactions on Information Theory}, vol.~59, no.~7, pp.
  4106--4128, July 2013.

\bibitem{Hao_Rasouli_Clerckx}
C.~Hao, B.~Rassouli, and B.~Clerckx, ``Degrees-of-freedom region of
  {MISO-OFDMA} broadcast channel with imperfect {CSIT},''
  \emph{arXiv:1310.6669}, October 2013.

\bibitem{Liu_Viswanath}
T.~Liu and P.~Viswanath, ``An extremal inequality motivated by multiterminal
  information-theoretic problems,'' \emph{IEEE Transactions on Information
  Theory}, vol.~53, no.~5, pp. 1839 -- 1851, May 2007.

\bibitem{Korner_Marton_images}
J.~Korner and K.~Marton, ``Images of a set via two different channels and their
  role in multiuser communication,'' \emph{IEEE Trans. Inform. Theory},
  vol.~23, pp. 751--761, Nov. 1977.

\bibitem{Arash_Jafar_GC15}
A.~G. Davoodi and S.~Jafar, ``Transmitter cooperation under finite precision
  {CSIT}: A {GDoF} perspective,'' \emph{Globecom}, Dec. 2015.

\bibitem{Jafar_Shamai}
S.~Jafar and S.~Shamai, ``Degrees of freedom region for the {MIMO} {X}
  channel,'' \emph{IEEE Trans. on Information Theory}, vol.~54, no.~1, pp.
  151--170, Jan. 2008.

\bibitem{Huang_Cadambe_Jafar}
C.~Huang, V.~Cadambe, and S.~Jafar, ``Interference alignment and the
  generalized degrees of freedom of the {X} channel,'' \emph{IEEE Transactions
  on Information Theory}, vol.~58, no.~8, pp. 5130--5150, August 2012.

\bibitem{Niesen_Maddah_Ali_X}
U.~Niesen and M.~A. Maddah-Ali, ``Interference alignment: From
  degrees-of-freedom to constant-gap capacity approximations,'' \emph{IEEE
  Transactions on Information Theory}, vol.~59, no.~8, pp. 4855 -- 4888, August
  2013.

\bibitem{Maric_Dabora_Goldsmith}
A.~G. I.~Maric, R.~Dabora, ``Relaying in the presence of interference:
  Achievable rates, interference forwarding and outer bounds,'' \emph{IEEE
  Transactions on Information Theory}, no.~7, July 2012.

\bibitem{Joudeh_Clerckx}
H.~Joudeh and B.~Clerckx, ``Sum rate maximization for {MU-MISO} with partial
  {CSIT} using joint multicasting and broadcasting,'' \emph{IEEE ICC}, 2015.

\bibitem{Geng_TIN_opt}
C.~Geng, N.~Naderializadeh, S.~Avestimehr, and S.~Jafar, ``On the optimality of
  treating interference as noise,'' \emph{IEEE Trans. on Information Theory},
  vol.~61, no.~4, pp. 1753--1767, Apr. 2015.

\bibitem{Jafar_TIM}
S.~A. Jafar, ``{Topological Interference Management through Index Coding},''
  \emph{IEEE Trans. on Information Theory}, no.~1, pp. 529--568, Jan. 2014.

\bibitem{Naderi_Avestimehr}
N.~Naderializadeh and A.~S. Avestimehr, ``Interference networks with no csit:
  Impact of topology,'' \emph{ArXiv}, vol. abs/1302.0296, 2013.

\bibitem{Jafar_Vishwanath_GDoF}
S.~Jafar and S.~Vishwanath, ``{Generalized Degrees of Freedom of the Symmetric
  Gaussian K User Interference Channel},'' \emph{IEEE Transactions on
  Information Theory}, vol.~56, no.~7, pp. 3297--3303, July 2010.

\end{thebibliography}
\end{document}